\documentstyle[12pt,epsfig]{article}
\newcommand{\be}{\begin{equation}}
\newcommand{\bea}{\begin{eqnarray}}
\newcommand{\eea}{\end{eqnarray}}
\newcommand{\ba}{\begin{array}}
\newcommand{\ea}{\end{array}}
\newcommand{\ee}{\end{equation}}

\newcommand{\nonu}{\nonumber \\[.5mm]}
\newcommand{\A}{&\!\!\!}
\newcommand{\slS}{{\cal S} \!\!\! \raisebox{0.2ex}{/}}
\expandafter\ifx\csname mathbbm\endcsname\relax

\else Kim

\fi
\def\appendix{{\newpage\section*{Appendix}}\let\appendix\section%
        {\setcounter{section}{0}
        \gdef\thesection{\Alph{section}}}\section}
\begin{document}
\begin{titlepage}
\hfill
\vbox{
    \halign{#\hfil         \cr
           CERN-TH/2001-123 \cr
           hep-th/0105170  \cr
           } 
      }  
\vspace*{30mm}
\begin{center}
{\Large {\bf The String Dual of an ${\cal N}=(4,0)$ 
Two-Dimensional Gauge Theory}\\} 

\vspace*{15mm}
\vspace*{1mm}
Madoka Nishimura

\vspace*{1cm} 

{\it Theory Division, CERN \\
CH-1211, Geneva  23, Switzerland}\\

\vspace*{.5cm}
\end{center}

\begin{abstract}
We find solutions of the six-dimensional maximal supergravity 
by adding a perturbation of vector fields to the solution AdS${}_3$ 
$\times$ S${}^3$. For certain perturbations the solution represents 
a dual description of an ${\cal N}=(4,0)$ field theory in two 
dimensions by the AdS/CFT correspondence. 
\end{abstract}
\vskip 4cm

May 2001

\end{titlepage}

\newpage
\section{Introduction}
%
One of the approaches to find field theories with a reduced 
supersymmetry in the context of the AdS/CFT correspondence 
\cite{MAL}--\cite{WITTEN} (for a review see 
ref.\ \cite{AGMOO},) is to use the brane 
polarization \cite{MYERS}.  D$p$-branes are polarized by 
background antisymmetric tensor fields, 
even with the form degree larger than $p+1$. 
As an extension of the solution AdS${}_5$ $\times$ S${}^5$ 
in the ten-dimensional type IIB supergravity, the authors of 
ref.\ \cite{PS} put a perturbation to it and found a new solution. 
By invoking the AdS/CFT correspondence, they argued that this solution 
is a dual of the ${\cal N}=4$ super Yang--Mills theory with a mass 
perturbation, which breaks the supersymmetry to ${\cal N}=1$.
\par
In the ten-dimensional supergravity 
we try to find a large-$N$ $p$-brane solution with a perturbation of 
electric RR $(p+4)$-form field strengths or its dual magnetic 
RR $(10-(p+4))$-form field strengths. Since the geometry of the 
large-$N$ limit of $p$-brane solutions contains AdS${}_{p+2}$, 
we perturb the AdS space with the $(6-p)$-form field strengths.
In the D3-brane case \cite{PS} the RR 3-form field strengths are put 
in the directions of S${}^5$, which is the partner of the AdS space. 
In the AdS/CFT correspondence the symmetry of the sphere is 
related to the R-symmetry of the field theory. 
When the R-symmetry is broken by the perturbation on the sphere, 
we have a field theory with a reduced supersymmetry. 
We expect that the rank of the field strengths for the perturbation 
should be half of the number of the sphere coordinates 
and the radial coordinate, that is ${1 \over 2}(10-(p+2)+1)$. 
The field strengths have a background within the 
($10-(p+2)+1$)-dimensional space S${}^{10-(p+2)}$ $\times$ ${\bf R}^+$. 
In the case of the eleven-dimensional supergravity compactified on 
AdS${}_4$ $\times$ S${}^7$ the branes are membranes 
($p=2$) and we perturb the solution by the 4-form field strengths 
within S${}^7$ $\times$ ${\bf R}^+$. 
\par
In this paper we consider a perturbation to the solution 
AdS${}_3$ $\times$ S${}^3$ in the maximal supergravity in six 
dimensions \cite{TANII}, which is obtained from the ten-dimensional 
supergravity by a compactification on T${}^4$. 
This solution is the geometry of the D1--D5 system. 
We put a perturbation of the RR field strengths of rank $4/2=2$ in 
S${}^3$ and see how many supersymmetries are preserved 
by the perturbation. We find a perturbation that breaks 
half of the supersymmetries and gives a chiral theory in two 
dimensions. According to the original idea of the dielectric 
perturbation, the rank of the dual magnetic RR field strength 
in ten dimensions should be $10-(1+4) = 5$. 
So let us consider that the rank 2 of the 5-form field 
strengths is in the sphere directions. 
We will discuss this perturbation in the supergravity
in the last section. 
\par
The RR fields in the type IIB supergravity are a scalar $\phi$, 
a rank-2 antisymmetric tensor $B_{MN}$ and a rank-4 $B_{MNPQ}$, 
where $M, N, \cdots = 0, 1, \cdots, 9$ are ten-dimensional world 
indices. If we compactify it on T${}^4$, we have the six-dimensional 
maximal Poincar\'e supergravity \cite{TANII}. Let us focus on the 
4-form field, which plays the role of perturbation field 
in the procedure. 
We should choose $B_{iIJK}$ for the field of the perturbation, 
where $i$ denotes the radial coordinate of AdS${}_3$ or a 
coordinate of S${}^3$, and $I,J,K$ denote coordinates of T${}^4$. 
In other words, the 5-form field strength in ten dimensions 
for the perturbation appears as a 2-form field strength in 
the compactified six-dimensional theory. 
\par
The organization of this paper is as follows. 
In sect.\ 2 we give the solution AdS${}_3$ $\times$ S${}^3$ 
in six-dimensional supergravity. 
Then, we introduce a perturbation of gauge fields 
and find a solution for the field equations; we 
see how many supersymmetries are preserved by this 
perturbation in sect.\ 3. Section 4 is devoted to discussions 
about this procedure and our conclusion. 
%
\section{Compactification to three-dimensional AdS}
%
Here we consider a solution for AdS${}_3$ $\times$ S${}^3$ in 
the six-dimensional Poincar\'e supergravity with the maximal 
supersymmetry \cite{TANII}. This six-dimensional theory is 
obtained from the ten-dimensional type IIB supergravity 
\cite{SCHWARZ}, \cite{HW} by a compactification on T${}^4$. 
\par
The six-dimensional maximal supergravity has a rigid SO(5,5) 
symmetry and a local SO(5) $\times$ SO(5) symmetry. 
The field content of the theory is a sechsbein $e_M{}^A$, five 
antisymmetric tensor fields $B^m_{MN}$, 16 vector fields 
$A_M^{\tilde{\mu}\dot{\tilde{\mu}}}$, 25 scalar fields 
$\phi_{\tilde{\mu}\dot{\tilde{\mu}}}^{\alpha\dot{\alpha}}$, 
eight Rarita--Schwinger fields $\psi_{+\alpha M}$, 
$\psi_{-\dot{\alpha}M}$ and 40 spinor fields 
$\chi_{+a\dot{\alpha}}$, $\chi_{-\dot{a}\alpha}$, where 
the signs on the spinor fields denote the chiralities. 
The indices take values $m, a = 1, \cdots, 5$ and 
$\tilde\mu, \dot{\tilde\mu}, \alpha, \dot\alpha = 1, \cdots, 4$. 
The indices $\tilde\mu\dot{\tilde\mu}$ represent a spinor index 
of SO(5,5), while $a, \dot{a}$ and $\alpha, \dot\alpha$ represent 
vector and spinor indices of SO(5) $\times$ SO(5), 
respectively. The representation of each field for SO(5,5) and 
SO(5) $\times$ SO(5) is given in Table 1. 
The field strengths of the antisymmetric tensor fields 
\be
H_{MNP}^m = 3 \partial_{[M} B_{NP]}^m 
\ee
and their duals $G_{MNP}^m$ belong to {\bf 10} of SO(5,5). 
\begin{center}
\arrayrulewidth=0.8pt
\def\arraystretch{1.2}
 \begin{tabular}{lcc} 
  \multicolumn{1}{c}{Fields} & SO(5,5) & SO(5)$\times$ SO(5) \\ \hline
  $e_M{}^A$ & {\bf 1} & {\bf 1} \\
  ($H_{MNP}^m,G_{MNP}^m$)& {\bf 10} & {\bf 1} \\
  $A_M^{\tilde{\mu}\dot{\tilde{\mu}}}$ & {\bf 16} & {\bf 1} \\
  $\phi_{\tilde{\mu}\dot{\tilde{\mu}}}^{\alpha\dot{\alpha}}$ 
  & {\bf 16} & ({\bf 4},{\bf 4})
  \\
  $\psi_{+\alpha M}$ & {\bf 1} & ({\bf 4},{\bf 1})\\
  $\psi_{-\dot{\alpha}M}$ & {\bf 1} & ({\bf 1},{\bf 4})\\
  $\chi_{+a\dot{\alpha}}$ & {\bf 1} & ({\bf 5},{\bf 4})\\
  $\chi_{-\dot{a}\alpha}$ & {\bf 1} & ({\bf 4},{\bf 5}) \\ \hline
 \end{tabular}
\end{center}
\par
We are only interested in the fields $e_M{}^A$, $B_{MN}^m$ and 
$A_M^{\tilde\mu\dot{\tilde\mu}}$ and set other fields to zero except 
$\phi_{\tilde{\mu}\dot{\tilde{\mu}}}^{\alpha\dot{\alpha}} 
= \delta^\alpha_{\tilde\mu} \delta^{\dot\alpha}_{\dot{\tilde\mu}}$. 
By this background of the scalar fields the indices $\tilde\mu$ 
and $\alpha$, $\dot{\tilde\mu}$ and $\dot\alpha$, and $m$ and $a$ 
are identified. The relevant part of the Lagrangian is 
\bea
{\cal L} \A = \A {1 \over 4} e_6 R 
- {1\over 12} e_6 H^a_{MNP} H^{aMNP} 
- {1 \over 4} e_6 G_{MN}^{\alpha\dot\alpha} 
G^{MN}_{\alpha\dot\alpha} \nonu
\A\A 
+ {1 \over 2} e_6 G_{MN}^{\alpha\dot\alpha} 
A_{P\beta\dot\alpha} 
(\gamma^a)_{\alpha}{}^{\beta} H_+^{aMNP} \nonu
\A\A
+ {1 \over 2} e_6 G_{MN}^{\alpha\dot\alpha} 
A_{P\alpha\dot\beta} 
(\gamma^{\dot{a}})_{\dot\alpha}{}^{\dot\beta} H_-^{\dot{a}MNP}, 
\label{lagrangian}
\eea
where $e_6 = |\det e_M{}^A|$ and $\gamma^a$ are SO(5) gamma 
matrices. $H_{\pm MNP}^a$ denote the self-dual and anti-self-dual 
parts of $H_{MNP}^a$: 
\be
H_{\pm MNP}^a = {1 \over 2}
\left( H^a \pm *_6 H^a \right)_{MNP}, 
\ee
where $(*_6 H^a)^{MNP}$ is the dual of $H^{a}_{MNP}$: 
\be
(*_6 H^a)^{MNP} = {1 \over 6} e_6^{-1} 
\epsilon^{MNPQRS} H_{QRS}^a. 
\ee
In this section we consider a unperturbed solution and 
set $A_M = 0$. 
The Einstein equation derived from eq.\ (\ref{lagrangian}) is 
\be
R_{MN} - {1 \over 2}g_{MN}R 
= H^a_{MPQ} H^a_{N}{}^{PQ} - {1 \over 6} g_{MN} H^a_{PQR}H^{aPQR}. 
\label{einstein}
\ee
By multiplying $g^{MN}$ we see that the scalar curvature vanishes. 
The field equations and the Bianchi identities of $B_{MN}^m$ 
are written as 
\be
\partial_M(e_6 H^{aMNP})= 0,
\qquad
\partial_M(e_6 (*_6 H^a)^{MNP})=0. 
\label{fieldeqofb}
\ee
\par
The metric for the compactification AdS${}_3 \times$ S${}^3$ is 
\bea
d s^2 
\A =  \A Z(r)^{-1} dx^\mu dx^\nu \eta_{\mu\nu} 
+ Z(r) dx^i dx^j \delta_{ij} \nonu
\A = \A Z(r)^{-1} dx^\mu dx^\nu \eta_{\mu\nu} 
+ Z(r) dr^2 + R^2 d \Omega_3^2, 
\label{metric3} 
\eea
where the indices take values $\mu, \nu, \cdots = 0, 1$ and 
$i, j, \cdots = 1, 2, 3, 4$. The coordinates of the AdS space are 
$x^\mu$ and $r = (x^i x^i)^{1 \over 2}$ and $d\Omega_3^2$ denotes the 
metric of a unit 3-sphere. $Z(r) = R^2 / r^2$ is a harmonic 
function on ${\bf R}^4$, where $R$ is the radius of the AdS 
space and S${}^3$. In terms of the 
parameters of string theory, it can be written as 
\be
R^2  = \sqrt{Q_1 Q_5} g_6 \alpha', 
\ee
where $Q_1$ and $Q_5$ are the charges of the D1- and D5-branes 
respectively, $g_6$ is the six-dimensional coupling constant, 
and $\alpha'$ is the Regge slope parameter.   
We take the following ansatz for the 3-form field strengths: 
\be
H_{\mu\nu i}^a = R^{-2} {\cal S}^a \epsilon_{\mu\nu} x_i,  \qquad
H_{ijk}^a = - r^{-4} R^2 {\cal T}^a \epsilon_{ijkl} x^l, \qquad
\rm{otherwise} = 0, 
\label{solof3form}
\ee
where ${\cal S}^a$ and ${\cal T}^a$ are some constants. 
We find that eqs.\ (\ref{metric3}), (\ref{solof3form}) are 
solutions of the field equations (\ref{einstein}), 
(\ref{fieldeqofb}) when ${\cal S}^a$, ${\cal T}^a$ satisfy 
\be
{\cal S}^a {\cal S}^a + {\cal T}^a {\cal T}^a = 2. 
\ee
\par
The next task is to see how many supersymmetries are preserved 
by this solution. The local supersymmetry transformations of 
the spinor fields are \cite{TANII}
\bea
\delta \psi_{+M\alpha} 
& = & \hat{\cal D}_M \epsilon_{+\alpha}
-{1 \over 4} H_{+MNP}^a (\gamma_a)_\alpha{}^\beta 
\Gamma^{NP} \epsilon_{+\beta}, \nonu
\delta \psi_{-M\dot{\alpha}} 
& = & \hat{\cal D}_M \epsilon_{-\dot{\alpha}}
-{1 \over 4} H_{-MNP}^{\dot{a}} 
(\gamma_{\dot{a}})_{\dot{\alpha}}{}^{\dot{\beta}} 
\Gamma^{NP} \epsilon_{-\dot{\beta}}, \nonu
\delta \chi_{+ a\dot{\alpha}} 
& = & {1 \over 12} H^a_{+MNP} \Gamma^{MNP} 
\epsilon_{-\dot{\alpha}}, \nonu
\delta \chi_{- \dot{a}\alpha} 
& = & {1 \over 12} H^{\dot a}_{-MNP} \Gamma^{MNP} 
\epsilon_{+\alpha}. 
\label{supersym}
\eea
We decompose the six-dimensional gamma matrices as 
\bea
\Gamma^A & = & \hat{\gamma}^A \otimes \bar{\gamma}_{\rm{4D}}
\qquad (\mbox{for } A=0,1),
\nonu
\Gamma^A & = & 1 \otimes \hat{\gamma}^A 
\qquad (\mbox{for } A=2,\cdots,5), 
\eea
where $\hat\gamma^A$ ($A=0,1$) and $\hat\gamma^A$ ($A=2, \cdots, 5$) 
are gamma matrices of SO(1,1) and SO(4) respectively, and 
\be
\bar{\gamma}_{\rm{2D}}  \equiv \hat{\gamma}^0 \hat{\gamma}^1, 
\qquad
\bar{\gamma}_{\rm{4D}} \equiv \hat{\gamma}^2 \cdots \hat{\gamma}^5.
\ee
Inserting the solution (\ref{solof3form}) and (\ref{metric3}) 
into (\ref{supersym}) the conditions $\delta \chi_\pm = 0$ require 
\be
-{1 \over 2r} x_i \bar{\gamma}_{\rm{2D}} \hat{\gamma}^i
\left({\cal S}^a \pm {\cal T}^a \right) \epsilon_{\mp} = 0. 
\ee
These equations are satisfied if we take $\epsilon_-=0$ and 
${\cal S}^a={\cal T}^a$, or $\epsilon_+=0$ and 
${\cal S}^a=-{\cal T}^a$. We choose the case $\epsilon_-=0$ and 
${\cal S}^a={\cal T}^a$. In this case the antisymmetric tensor 
fields are self-dual $*_6 H^a_{MNP} = H^a_{MNP}$. 
Finally, we have to show the existence of a solution for 
$\delta \psi_{+M} = 0$ in eq.\ (\ref{supersym}). 
We have checked that the integrability conditions 
for these equations 
\bea
& & \left[\hat{D}_\mu-{1 \over 2} {\cal S}^a (\gamma^a)
\hat{\gamma}_\mu \bar{\gamma}_{\rm 2D}\otimes{\hat\gamma}_i x^i
\bar{\gamma}_{\rm 4D} , 
\hat{D}_\nu+{1 \over 4} {\cal S}^b (\gamma^b)
\hat{\gamma}_\nu \bar{\gamma}_{\rm 2D}\otimes{\hat\gamma}_j x^j
\bar{\gamma}_{\rm 4D} 
\right] =0, 
\nonu & & 
\biggl[\hat{D}_\mu-{1 \over 2}{\cal S}^a (\gamma^a)
\hat{\gamma}_\mu \hat{\gamma}_{\rm 2D}\otimes{\hat\gamma}_l x^l
\bar{\gamma}_{\rm 4D}, 
\nonu & & \qquad\qquad
\hat{D}_i -r^{-2} {\cal S}^b (\gamma^b) \bar{\gamma}_{\rm 2D} 
\left\{1 \otimes 1 x_i
+ \bar{\gamma}_{\rm 2D}\otimes \bar{\gamma}_{\rm 4D} \hat{\gamma}_{ik}
x^k \right\}
\biggr]=0, 
\nonu
& & 
\biggl[\hat{D}_i -r^{-2} {\cal S}^a (\gamma^a) \bar{\gamma}_{\rm 2D} 
\left\{1 \otimes 1 x_i
+ \bar{\gamma}_{\rm 2D}\otimes \bar{\gamma}_{\rm 4D} \hat{\gamma}_{ik}
x^k \right\}, 
\nonu & & \qquad \qquad
\hat{D}_j -r^{-2} {\cal S}^b (\gamma^b) \bar{\gamma}_{\rm 2D} 
\left\{1 \otimes 1 x_j
+ \bar{\gamma}_{\rm 2D}\otimes \bar{\gamma}_{\rm 4D} \hat{\gamma}_{jl}
x^l\right\}
\biggr]=0 
\eea 
are indeed satisfied by the above solution. 
Thus, we find that half of the supersymmetry corresponding to 
the parameters $\epsilon_{+\alpha}$ are preserved. 
This solution corresponds to an ${\cal N} = (4,4)$ conformal field 
theory in two dimensions \cite{MAL}, \cite{BOER}. 
%
\section{Three-dimensional AdS supergravity with perturbation}
%
We consider a perturbation of the vector fields to the solution 
in Sect. 2. The linearized field equations of the vector fields in 
the background (\ref{metric3}), (\ref{solof3form}) are 
\be
d *_6 G_2^{\alpha\dot\alpha} 
+ 2 G_2^{\beta\dot\alpha} 
\wedge H_3^a (\gamma^a)_{\beta}{}^{\alpha} = 0, 
\qquad 
d G^{\alpha\dot\alpha}_2 = 0, 
\label{eqofmotionG}
\ee
where $G^{\alpha\dot\alpha}_{MN}$ is the field strengths 
the vector fields $A^{\alpha\dot\alpha}_M$ and we have used 
the self-duality of $H^a_{MNP}$. 
The six-dimensional Hodge dual $*_6$ here can be expressed by 
the four-dimensional flat one $*_4$ as 
\be
*_6 G_2 = Z^{-1} *_4 G_2 \wedge dx^0 \wedge dx^1. 
\ee
By using eqs.\ (\ref{metric3}) and (\ref{solof3form}) the first 
equation of eq.\ (\ref{eqofmotionG}) can be rewritten as 
\be
d \left[ Z^{-1} \left( *_4 G^{\alpha\dot\alpha}_2 
+ G^{\beta\dot\alpha}_2 
(\slS)_{\beta}{}^{\alpha} \right) \right] = 0, 
\label{eqofmotionG2}
\ee
where $\slS = {\cal S}^a \gamma^a$. 
Since ${\cal S}^a {\cal S}^a = 1$, we have $\slS{}^2 = 1$. 
\par
To solve these equations we introduce 2-forms with constant 
components $T^{\alpha\dot\alpha}_2 
= {1 \over 2} T^{\alpha\dot\alpha}_{ij} dx^i \wedge dx^j$, 
which satisfy the (anti)self-duality condition 
\be
*_4 T^{\alpha\dot\alpha}_2 
= \pm T^{\alpha\dot\alpha}_2. 
\label{dual3}
\ee
Their explicit forms are 
\be
T^{\alpha\dot\alpha} _2 
= m^{\alpha\dot\alpha}  dz^1 \wedge d\bar{z}^2 
+ m^{\alpha\dot\alpha}  d\bar{z}^1 \wedge dz^2 
\ee
for the self-dual case and 
\be
T^{\alpha\dot\alpha} _2 
= m^{\alpha\dot\alpha}  dz^1 \wedge dz^2 
+ m^{\alpha\dot\alpha} d\bar{z}^1 \wedge d\bar{z}^2 
\ee
for the antiself-dual case, where $m^{\alpha\dot\alpha}$ 
is a constant matrix and $z^1, z^2$ are complex coordinates 
given by 
\be
z^1={x^2+ix^4 \over \sqrt{2}}, \qquad 
z^2={x^3+ix^5 \over \sqrt{2}}. 
\ee
In the following we suppress the indices $\alpha\dot\alpha$, 
for simplicity unless necessary. 
Let us define 
\be
V_{ij} = {x^k \over r^2} (x^iT_{kj} + x^jT_{ik}), \qquad
S_i = T_{ij} x^j. 
\ee
We can show the following relations: 
\bea
dS_1 \A = \A 2 T_2, \qquad
dT_2 = 0, \qquad
dV_2 = -2d(\ln r) \wedge T_2, 
\nonu
V_2 \A = \A d(\ln  r) \wedge S_1, \qquad 
d(r^p S_1)  =   r^p(2T_2+pV_2) 
\eea
and
\be
*_4 V_2 = \pm ( T_2 - V_2 ). 
\label{dual4}
\ee
\par
Using these tensors we can construct a solution to the linearized 
field equations (\ref{eqofmotionG2}) as follows. We make an ansatz 
\be
G_2 = {1 \over 2} r^p (\alpha T_2 + \gamma V_2) (1 - \slS) 
      + {1 \over 2} r^q (\beta T_2 + \delta V_2) (1 + \slS). 
\ee
The Bianchi identity $d G_2 = 0$ in eq.\ (\ref{eqofmotionG}) gives 
\be
\gamma = {p \alpha \over 2} , \qquad 
\delta = {p \beta \over 2}, \qquad
G_2 = {1 \over 4} \alpha d(r^p S_1) (1-\slS) 
+ {1 \over 4} \beta d(r^q S_1) (1+\slS). 
\ee
Using the duality properties (\ref{dual3}) and (\ref{dual4}) we have 
\bea
*_4 G_2 - G_2 
\A = \A {1 \over 4} \alpha r^p \left[ (\pm p \pm 2 - 2) T_2 
+ (\mp p - p) V_2 \right] (1-\slS) \nonu
\A\A + {1 \over 4} \beta r^q \left[ (\pm q \pm 2 + 2) T_2 
+ (\mp q + q) V_2 \right] (1+\slS). 
\eea
The field equation in (\ref{eqofmotionG2}) then gives
\be
p^2 + 6p +(4 \mp 4)=0, \qquad q^2 + 6q + (4 \pm 4) = 0. 
\label{peq}
\ee
Let us first consider the $\alpha$ terms. 
For the lower sign, there are two solutions of eq.\ (\ref{peq}):
\bea
p \A = \A -4 ; \qquad G_2 
= {\alpha \over 2} r^{-4} (T_2 - 2V_2) (1-\slS), \nonu 
p \A = \A -2 ; \qquad G_2 
= {\alpha \over 2} r^{-2} (T_2 - V_2) (1-\slS).
\label{lowersol}
\eea
For the upper sign, there are also two solutions of 
eq.\ (\ref{peq}): 
\bea
p \A = \A -6 ; \qquad G_2 
= {\alpha \over 2} r^{-6} (T_2 - 3V_2) (1-\slS), 
\nonu 
p \A = \A 0 ; \qquad G_2 
= {\alpha \over 2} T_2 (1-\slS). 
\label{uppersol}
\eea
Similarly, the solutions for the $\beta$ terms are 
\bea
q \A = \A -6 ; \qquad G_2 
= {\beta \over 2} r^{-6} (T_2 - 3V_2) (1+\slS), 
\nonu 
q \A = \A 0 ; \qquad G_2 
= {\beta \over 2} T_2 (1+\slS) 
\label{lowersol2}
\eea
for the lower sign and 
\bea
q \A = \A -4 ; \qquad G_2 
= {\beta \over 2} r^{-4} (T_2 - 2V_2) (1+\slS), 
\nonu 
q \A = \A -2 ; \qquad G_2 
= {\beta \over 2} r^{-2} (T_2 -V_2) (1+\slS) 
\label{uppersol2}
\eea
for the upper sign. 
\par
Let us count how many supersymmetries are preserved by this 
perturbation. The additional terms to the supertransformations 
generated by the perturbation are 
\bea
\delta \psi_{-\mu\dot\alpha} 
\A = \A {1 \over 8}i G_{jk\beta\dot\alpha}
(\hat\gamma_\mu \times \hat\gamma^{jk}) 
\epsilon_+^\beta, \nonu
\delta \psi_{-i\dot\alpha} 
\A = \A -{3 \over 4}i G_{ij\beta\dot\alpha}
({\bf 1} \times \hat\gamma^j) \epsilon_+^\beta
+{1 \over 8}i G_{jk\beta\dot\alpha} 
({\bf 1} \times \hat\gamma_i{}^{jk}) 
\epsilon_+^\beta, \nonu
\delta \chi_{+a\dot{\alpha}} 
\A = \A {1 \over 4} G_{ij\beta\dot{\alpha}}
({\bf 1} \times \hat\gamma^{ij})
\epsilon_+^\alpha (\gamma_a)_\alpha{}^\beta. 
\eea
Generically, these terms do not vanish and supersymmetry is 
completely broken. However, for particular backgrounds 
$G_{ij\beta\dot{\alpha}}$ some of the supersymmetries are 
preserved. Let us consider the $\alpha$ terms with $p = -4$ 
for the lower sign or those with $p=0$ for the upper sign, 
in which $G_{ij\beta\dot{\alpha}}$ is self-dual. 
If we impose the condition 
$\bar\gamma_{4D} \epsilon_+ = \epsilon_+$, 
$\delta \psi_{-\mu\dot\alpha}$ and 
$\delta \chi_{+a\dot{\alpha}}$ are shown to vanish 
by using the identity $\hat\gamma^{ij} = - {1 \over 2} 
\epsilon^{ijkl} {\hat\gamma}_{kl} \bar\gamma_{4D}$. 
This condition does not guarantee the vanishing of 
$\delta \psi_{-i\dot\alpha}$. Since it is  proportional 
to $\epsilon^\alpha m_{\alpha\dot\alpha}$, 
the remaining supersymmetries are determined by zero 
eigenvalues of the matrix $m$. 
For instance, let us consider the following form of the matrix 
\be
m_{\alpha\dot\alpha} = {1 \over 2} m^{mn} 
(\gamma_{mn})_{\alpha\dot\alpha}. 
\ee
Here, $\gamma_{mn}$ is an antisymmetrized product of gamma 
matrices for the SO(4) subgroup in SO(5), which is chosen 
such that the chirality matrix of SO(4) is $\slS$. Then, 
$\delta \psi_{-i\dot\alpha}$ vanishes when $m_{mn}$ 
is self-dual. Since $\bar\gamma_{4D} \epsilon_+ = \epsilon_+$ 
means $\bar\gamma_{2D} \epsilon_+ = \epsilon_+$ for 
six-dimensional spinors $\epsilon_+$ of positive chirality, 
${\cal N} = (4,0)$ supersymmetry in two dimensions is 
preserved in this case. 
%
\section{Summary and Conclusion}
%
We found a supergravity solution with the reduced supersymmetry, 
which is the dual string theory of an ${\cal N}=(4,0)$ field theory. 
We discussed the following. 
\par
First, we constructed an ${\cal N}=(4,0)$ field theory from 
the point of view of supergravity. We should also discuss 
it from the field theory side. 
We need to find an operator  
corresponding to the perturbation in our solution. 
{}From the $r$-dependence of the perturbation 
we can read off the conformal dimension of the operator. 
It is also an interesting problem to find out a holographic 
renormalization group flow from ${\cal N}=(4,4)$ to ${\cal N}=(4,0)$ 
in two-dimensional field theory in the context of the supergravity. 
\par
Second, we are interested in how can be interpreted the puffing up 
of the D-branes in S${}^3$ $\times$ T${}^4$. Since we did not put 
the antisymmetric tensor in the whole space of the sphere S${}^3$, 
the polarization does not seem to take place. But if we consider 
T${}^4$ in addition to the sphere, we will see the polarization, 
but we need to consider the topology there. 
\par
Third, we considered the gauge field potentials, which are 
related to the D1-branes. But we did not consider the precise 
effects of the D5-branes when we discussed the perturbation. 
In our work, the hidden three directions belong to the four 
directions in the compactified manifold T${}^4$, where the D5-branes 
also exist. The torus T${}^4$ should be treated carefully. 
At least in the six-dimensional AdS${}_3$ $\times$ S${}^3$ space, 
we see the RR charge is in S${}^1$, just like the 2-forms are 
in S${}^2$ for the case of large-$N$ D3-branes \cite{PS}. 
\par
\bigskip
%
\begin{flushleft} 
\large{\bf{Acknowledgements}} 
\end{flushleft} 
The author would like to thank T. Eguchi and Y. Tanii for 
useful discussions and 
Theory Division of CERN for hospitality. 
This work is supported in part by the Swiss National Science 
Foundation and the Japan Society for the Promotion of Science. 
%
%
\newcommand{\NP}[1]{{\it Nucl.\ Phys.\ }{\bf #1}} 
\newcommand{\PL}[1]{{\it Phys.\ Lett.\ }{\bf #1}} 
\newcommand{\CMP}[1]{{\it Commun.\ Math.\ Phys.\ }{\bf #1}} 
\newcommand{\MPL}[1]{{\it Mod.\ Phys.\ Lett.\ }{\bf #1}} 
\newcommand{\IJMP}[1]{{\it Int.\ J. Mod.\ Phys.\ }{\bf #1}} 
\newcommand{\PRe}[1]{{\it Phys.\ Rev.\ }{\bf #1}} 
\newcommand{\PRL}[1]{{\it Phys.\ Rev.\ Lett.\ }{\bf #1}} 
\newcommand{\AP}[1]{{\it Ann.\ Phys.\ }{\bf #1}} 
\newcommand{\ATMP}[1]{{\it Adv.\ Theor.\ Math.\ Phys.\ }{\bf #1}} 

\end{document}